\documentstyle[12pt]{article}

\parskip=\medskipamount                 
\def\beee{\begin{equation}}
\def\eeee{\end{equation}}
\def\dggg{^{\dagger}}
\begin{document}
\bibliographystyle{unsrt}

\begin{center}
{\Large \bf CONSERVATION OF STATISTICS AND GENERALIZED GRASSMANN
NUMBERS}\\[7mm]
O.W. Greenberg\footnote{Supported in
part by a Semester Research Grant from the University of Maryland, College
Park and by the National Science Foundation.\\
e-mail address: greenberg@umdhep.umd.edu}

{\it Center for Theoretical Physics\\
Department of Physics and Astronomy\\
University of Maryland\\
College Park, MD~~20742-4111}\\[5mm]
Preprint number 96-3\\
\end{center}

\vspace{2mm}

\begin{center}
{\bf Abstract}
\end{center}
Conservation of statistics requires that fermions be coupled to Grassmann
external sources.  Correspondingly, conservation of statistics requires that
parabosons, parafermions and quons be coupled to external sources that are
the appropriate generalizations of Grassmann numbers.

\newpage

{\bf 1. Introduction}

The classic constraints of conservation of statistics in theories
with bosons and fermions are that all
terms in the Hamiltonian must have an even number of Fermi fields, and
composites of bosons and fermions are bosons, unless they contain
an odd number of fermions, in which case they are fermions.  Thus an even
number of fermions must
participate in any reaction and no reaction can involve only
one fermion.

With the introduction of new kinds of particle statistics, such as parabosons
and parafermions\cite{hs,gm} and
quons\cite{pr43,g1,g2,physica,fi,bo,speicher,zag}
it is relevant to consider possible Hamiltonian densities, including
couplings to external sources, that involve fields obeying the new statistics,
rather than the familiar Bose and Fermi statistics.  One is tempted to carry
over constructions used for Bose and Fermi fields to the new types of fields.
The purpose of this paper is to point out that care must be exercised to
ensure that the Hamiltonian density is an effective Bose operator in the sense
that
\begin{equation}
[{\cal H}(x),\phi(y)]=0,~ |{\bf x}-{\bf y}| \rightarrow \infty.
\label{eq:com}
\end{equation}
for {\it all} fields $\phi$, regardless of whether $\phi$ is Bose, Fermi,
parabose, parafermi or quon.
This requirement, which is necessary in order that the
energy of widely separately
particles is the sum of the energies of the individual particles, leads to
the conservation of statistics discussed above.
Sudbery\cite{sud} pointed out the implications of this constraint for
particles with anomalous statistics.  In the case of couplings to external
sources where the particle number is not conserved, the additivity of energy
requirement is replaced by additivity of transition matrix elements.
The simplest extension of conservation of statistics is that a
single parabose, parafermi or quon particle cannot couple to ``normal''
(Bose or Fermi) particles.  To couple these ``anomalous'' particles
to external sources, I introduce parabose, parafermi
and quon analogs of Grassmann numbers.  Their external sources must be coupled
to the quantized fields in such a way that the term in the Hamiltonian is an
effective Bose operator; otherwise additivity of transition matrix elements
for widely separated
subsystems would be violated.  Since qualitative issues concerning
statistics should be the same for noninteracting particles as for particles
whose interactions vanish for large space separation, I give the discussion
in terms of noninteracting particles.  In this case, using discrete notation,
the condition for an effective Bose operator is
$[n_i,a\dggg_j]_-=\delta_{ij}a\dggg_j$ without external sources and
$[s_i,a\dggg_j]_-=\delta_{ij}g^{\star}_j$ with an external source, where
$s_i$ is the external source term and $g^{\star}_j$ is the appropriate
generalization of a Grassmann number.

As an example
of what can go wrong, consider a collection of free identical
Fermi particles with annihilation and creation operators,
$a_i~{\rm and}~a_i\dggg$,
labelled by quantum numbers $i$.  In all cases, we want an external source to
contribute equally to each of these particles.  We must couple the
external source to the Fermi particles using Grassmann numbers, $f_i$,
that obey $[f_i,f^{\star}_j]_+=0$, and take
as the external Hamiltonian,
\beee
H_{ext}=\sum_i(f_i^{\star}a_i+a\dggg_if_i).
\eeee
Then,
\beee
[H_{ext},a\dggg_k]_-=f_k^{\star},                        \label{fstar}
\eeee
and, acting on a state of several fermions,
\beee
H_{ext}a\dggg_1 a\dggg_2 \cdots a\dggg_n|0\rangle=(f_1^{\star} a\dggg_2
\cdots a\dggg_n + a\dggg_1 f_2^{\star} \cdots a\dggg_n + \cdots
a\dggg_1 a\dggg_2 \cdots f_n^{\star})|0\rangle.
\eeee
Here, each fermion is treated in an equivalent way by the external source.  If
we had not coupled the external source using Grassmann numbers, but instead
used
c-nos., $j_i$, then we would have had
\beee
H_{ext}=\sum_i(j_i^{\star}a_i+a\dggg_ij_i),~[H_{ext},a\dggg_k]_-=
j_k+2\sum_i(a\dggg_ij_ia\dggg_k-a\dggg_kj^{\star}_ia\dggg_i)
\eeee
and, acting on a state of several fermions,
\beee
H_{ext}a\dggg_1 a\dggg_2 \cdots a\dggg_n|0\rangle=(j_1^{\star} a\dggg_2
\cdots a\dggg_n - a\dggg_1 j_2^{\star} \cdots a\dggg_n + \cdots +(-1)^{n-1}
a\dggg_1 a\dggg_2 \cdots j_n^{\star})|0\rangle,
\eeee
so the interactions of the successive fermions with the external source
would have alternated in sign.  If one considers a transition matrix element
between a state with $n$ particles and a state with $n \pm 1$ particles, the
contribution to the transition matrix element from the Fermi particles adds in
the case in which the Fermi particles are coupled to the external sources with
Grassmann numbers, but the signs of the contributions from the Fermi particles
alternates in the case in which the particles are coupled with c-numbers.
Because equivalent particles should contribute in an equivalent way, the
external sources must be Grassmann numbers in this case.

An analogous issue
arises in considering the choice of Hamiltonian in a theory of noninteracting
quons.  The commutation relation for the quons is
\beee
a_i a_j\dggg -q a_j\dggg a_i=\delta_{ij};
\eeee
there is no relation that allows transposing two quon creation or two quon
annihilation operators\cite{pr43}.
Consider two possibilities:  (a) the number operator, Hamiltonian, etc., have
their usual algebraic form,
\beee
n_i=a\dggg_i a_i,~~H=\sum_i \omega_i a\dggg_i a_i,~~ {\rm etc.}
\eeee
or (b) the number operator, etc., have the usual commutators with the
annihilation and creation operators,
\beee
[n_i,a\dggg_j]_-=\delta_{ij}, ~~[H,a\dggg_i]_-=\omega_i, ~~
[{\bf P},a\dggg_i]_-={\bf p}_i a\dggg_i.
\eeee
For case (a), the energy equation for a state of $n$ identical quons is
\beee
H|a\dggg_1 a\dggg_2 \cdots a\dggg_n\rangle=
\sum_j q^{j-1} \omega_j|a\dggg_1 a\dggg_2 \cdots a\dggg_n\rangle.
\eeee
In this case, the identical noninteracting quons contribute to the energy
with different
powers of $q$ depending on where in the state vector they appear.  This is
unreasonable, since identical noninteracting particles should contribute to the
energy in an equivalent way.
Another problem with this choice
is that the algebra of the generators of space-time symmetry groups will not be
satisfied.  To see this, let the momentum operator be
\beee
{\bf P}= \sum_i {\bf p}_i a\dggg_i a_i.
\eeee
The commutator of these observables is
\beee
[H,P]_-=q\sum_{i,j}\omega_i {\bf p}_j(a\dggg_i a\dggg_j a_i a_j-
a\dggg_j a\dggg_i a_j a_i).
\eeee
For the Bose and Fermi cases the two terms cancel;
however, for the quon case there is
no commutation relation among annihilation or among creation operators and
these terms do not cancel.  Thus,
except for $q=0$,
the energy and momentum operators cannot obey the correct algebra in case (a).
In case (b),  construct $n_i$ so that
\beee
[n_i, a\dggg_j]_-=\delta_{ij} a\dggg_j.
\eeee
A straight-forward calculation shows that the energy and, in the external
source case, the
transition matrix elements of noninteracting
particles are additive, and that the space-time generators obey the correct
algebra.
I made this choice for the special case (the Cuntz algebra\cite{cun})
of $q=0$\cite{owg}, and also made this choice for general
$q$\cite{pr43}.
For the special case of $q=0$,
I found the exact expression for the number operator, from which
the space-time symmetry operators can be constructed.  In the latter case, I
gave the first few terms of the number operator;  the complete formula for the
number operator was given by Stanciu\cite{stan}.
I conclude that (b), choosing the annihilation and creation operators to have
the usual commutation relations with the number operator, is the correct
choice.

The corresponding error with external sources is to couple the quons to a
c-number external source $j_i$ using
\beee
H_{ext}=\sum_i(j_i^{\star}a_i+a\dggg_ij_i).
\eeee
Then, acting on a state of several quons,
\beee
H_{ext}a\dggg_1 a\dggg_2 \cdots a\dggg_n|0\rangle=(j_1^{\star} a\dggg_2
\cdots a\dggg_n + qa\dggg_1 j_2^{\star} \cdots a\dggg_n + \cdots
q^{n-1}a\dggg_1 a\dggg_2 \cdots j_n^{\star})|0\rangle.
\eeee
Here, the powers of $q$ replace the powers of $(-1)$ in the Fermi case
discussed
above.  The
contributions to transition matrix elements acquire corresponding factors of
powers of $q$.  The external sources must be quon analogs of Grassmann numbers
in order that the contributions to transition matrix elements of widely
separated quons be additive.  Because quons were coupled to external sources
with c-numbers in \cite{fivel}, the conclusions of that paper are not reliable.

A further problem with \cite{fivel} is that in Model 2 of this reference the
$q$-exponential is not unitary:  the unitary
evolution operator does not have the form $exp(-itH)$, with $H$ time
independent, but rather has this form with $H(t)$ having the time dependence
implied by the peculiarities of the $q$-exponential.  The repair of this
nonunitarity introduces an uncontrolled time dependence in the Hamiltonian.
Since the large-time
dependence of the occupation number is crucial, this uncontrolled time
dependence is a serious flaw.

What is true for the coupling of external sources to quons is also true
for the coupling of parabosons and of parafermions to external sources:  in all
cases, the coupling must involve the appropriate analog of Grassmann numbers
and
the external Hamiltonian must be an effective Bose operator.
The commutation relations for these Grassmann analogs do not seem to appear in
the literature.  I supply them below.

{\bf 2. Coupling to external sources for parabosons and parafermions}

Green's trilinear commutation relations for parabose and
parafermi operators are
\beee
[n_{kl}, a\dggg_m]_-=\delta_{lm}a\dggg_k,
\eeee
where
\begin{equation}
n_{kl}=\frac{1}{2}([a^{\dagger}_k,a_l]_{\pm} \mp p \delta_{kl}), \label{nbf}
\end{equation}
and the upper (lower) sign is for parabosons (parafermions).
Since Eq.(\ref{nbf}) is trilinear, two conditions are necessary to
fix the Fock-like representation:  the usual vacuum condition is
\begin{equation}
a_k|0\rangle=0;         \label{vacp}
\end{equation}
the new condition
\begin{equation}
a_ka^{\dagger}_l|0\rangle=p \delta_{kl}, ~~p~~ {\rm integer},   \label{1p}
\end{equation}
contains the integer $p$ that is the order of the parastatistics.
The Hamiltonian for free particles obeying parastatistics has the same form,
in terms of the number operators, as for Bose and Fermi statistics,
\begin{equation}
H=\sum_k \epsilon_k n_k,~ {\rm where,~ as~ usual}~
[H,a^{\dagger}_l]_-=\epsilon_l a^{\dagger}_l.   \label{pcr}
\eeee
For interactions with an external source, introduce para-Grassmann
numbers that make the interaction Hamiltonian an effective Bose operator.
Require
\beee
[H_{ext}, a^{\dagger}_l]=c^{\star}_l .   \label{pext}
\eeee
This is accomplished by choosing
\beee
H_{ext}=\sum_{k}\frac{1}{2}([c^{\star}_k, a_k]_{\pm}+
[a\dggg_k,c_k]_{\pm}),  \label{hpara}
\eeee
where the para-Grassmann numbers $c_k$ and $c\dggg_k$ obey
\beee
[[c^{\star}_k,c_l]_{\pm},c^{\star}_m]_-=0,~
[[c^{\star}_k,a_l]_{\pm},a\dggg_m]_-
=2 \delta_{lm}c^{\star}_k,~{\rm etc.},  \label{pgrass}
\eeee
and the upper (lower) sign is for parabose-Grassmann (parafermi-Grassmann)
numbers.  The ``etc.'' in Eq.(\ref{pgrass}) means that when some of the $c$'s
or $c\dggg$'s are replaced by an $a$ or an $a\dggg$, the relation retains its
form, except when the $a$ and $a\dggg$ can contract, in which case the
term with the contraction appears on the right-hand-side.

{\bf 3. Coupling to external sources for quons}

The case of quons differs from all the previous cases in that the
external source Hamiltonian
is of infinite degree, instead of being bilinear.  (The Hamiltonian for free
particles is also of infinite degree\cite{pr43}.)  Since the infinite series is
simple in the special case of $q=0$\cite{owg},
I discuss this case first.
In that case,
the commutation relation is
\beee
a_ka\dggg_l=\delta_{kl},   \label{5}
\eeee
with the usual vacuum condition, Eq.(\ref{vacp}).
To construct observables, we want number operators and transition operators
that obey
\beee
[n_k,a\dggg_l]_-=\delta_{kl}a\dggg_l,~~[n_{kl},a\dggg_m]_-=\delta_{lm}a\dggg_k.
 \label{62}
\eeee
Once Eq.(\ref{62}) holds, the Hamiltonian and other observables can be
constructed in the usual way; for example,
\beee
H=\sum_k \epsilon_k n_k,~~ {\rm etc.}  \label{8}
\eeee
The obvious thing is to try
\beee
n_k=a\dggg_k a_k.                                        \label{9}
\eeee
Then
\beee
[n_k,a\dggg_l]_-=a\dggg_ka_ka\dggg_l-a\dggg_la\dggg_ka_k.  \label{10}
\eeee
The first term in Eq.(\ref{10}) is $\delta_{kl}a\dggg_k$ as desired; however
the second term is extra and must be canceled.  This can be done by adding the
term $\sum_ta\dggg_ta\dggg_ka_ka_t$ to the term in Eq.(\ref{9}).  This cancels
the extra term, but adds a new extra term, that must be canceled by another
term.  This procedure yields an infinite series for the number operator
and for the transition operator,
\beee
n_{kl}=a\dggg_ka_l+\sum_ta\dggg_ta\dggg_ka_la_t+\sum_{t_1,t_2}a\dggg_{t_2}
a\dggg_{t_1}a\dggg_ka_la_{t_1}a_{t_2}+ \dots   \label{11}
\eeee
As in the Bose case, this infinite series for the transition or number
operator defines an unbounded operator whose domain includes states made by
polynomials in the creation operators acting on the vacuum.

The quon-Grassmann numbers must satisfy
\beee
c_k c^{\star}_l=0;~c_k a\dggg_l=0;~
a_kc^{\star}_l=0.              \label{820}
\eeee
Then $H_{ext}$ must be chosen to obey
\beee
[H_{ext},a\dggg_l]_-=c^{\star}_l.              \label{821}
\eeee
This is accomplished by choosing
\beee
H_{ext}=\sum_k
(c^{\star}_ka_k+a\dggg_kc_k)+\sum_k\sum_ta\dggg_t(c^{\star}_ka_k+a\dggg_kc_k)a_t
+ \cdots                                   \label{85}
\eeee
in analogy with Eq.(\ref{11}).

The general quon algebra\cite{pr43} is
\beee
a_k a^{\dagger}_l-q a^{\dagger}_l a_k=\delta_{kl},         \label{p}
\eeee
with the usual vacuum condition, Eq.(\ref{vacp}).
For observables without an
external source, one again needs a set of number operators $n_k$ such that
\beee
[n_k,a\dggg_l]_-=\delta_{kl}a\dggg_l.           \label{81}
\eeee
Like the $q=0$ case, the expression for $n_k$ or $n_{kl}$ is an infinite series
in creation and annihilation operators; unlike the $q=0$ case, the coefficients
are complicated.  The first two terms are
\beee
n_{kl}=a^{\dagger}_ka_l
+ (1-q^2)^{-1} \sum_t (a^{\dagger}_t a^{\dagger}_k
-q a^{\dagger}_ka^{\dagger}_t)(a_la_t-qa_ta_l) + \cdot \cdot \cdot. \label{221}
\eeee
Here I have given the transition number operator $n_{kl}$ for $k \rightarrow l$
since this takes no extra effort.  The general formula for the number operator
is given in \cite{stan} following a conjecture of Zagier \cite{zag}.
As before, the Hamiltonian is
\beee
H=\sum_k \epsilon_k n_k,~~{\rm with}~~[H,a\dggg_l]_-=\epsilon_la\dggg_l.
                \label{82}
\eeee
For an external source, we again require that $H_{ext}$ be an effective Bose
operator and again accomplish this using quon-Grassmann numbers.  Now these
obey
\beee
c_k c^{\star}_l-qc^{\star}_lc_k=0;~c_k a\dggg_l-qa\dggg_l c_k=0;~
a_kc^{\star}_l-qc^{\star}_la_k=0,              \label{xyz}
\eeee
and $H_{ext}$ obeys
\beee
[H_{ext},a\dggg_l]_-=c^{\star}_l.               \label{822}
\eeee
For this to work, we need
\begin{eqnarray}
H_{ext}=\sum_k(c^{\star}_ka_k+a\dggg_kc_k)+(1-q^2)^{-1}
\sum_t(a\dggg_tc^{\star}_k-q
c^{\star}_ka\dggg_t)(a_ka_t-qa_ta_k)     \nonumber          \\
+\sum_k(1-q^2)^{-1}(a\dggg_ta\dggg_k-qa\dggg_ka\dggg_t)(c_ka_t-qa_tc_k) +
\cdots
           \label{8002}
\end{eqnarray}
The general result for $H_{ext}$ can be gotten from the number operator of
Ref.\cite{stan} by replacing some of the $a$'s and $a\dggg$'s by $c$'s and
$c^{\star}$'s in analogy to the change from Eq.(\ref{221}) to Eq.(\ref{8002}).
If, instead, we incorrectly choose
$H_{ext}=\sum_k(j^{\star}_ka_k+a\dggg_kj_k)$,
where $j$ is a $c$-number, then the interactions of noninteracting systems
(or of widely separated subsystems) with the external sources are not
additive as illustrated in the introduction.
Because this point was not recognized, the bound on
laser intensities due to a small violation of
Bose statistics for photons claimed in \cite{fivel} cannot be taken seriously.

{\bf 4. Difficulties in obtaining high-precision bounds on Bose statistics}

There are two reasons that make it difficult to get high-precision bounds on
the
validity of Bose statistics for photons and other presumed bosons.  (1) Stable
matter is made of fermions, not bosons, so one cannot search for stable or
quasi-stable states of bosons that exhibit anomalous statistics, nor can one
search for transitions to such states.  (2) It is difficult to make a
high-precision measurement of deviations from the Bose distribution in
macroscopic samples, because the effect due to a possible small concentration
of anomalous states will be swamped by the much larger number of normal states.
This problem also arises in the case of fermions.  A general discussion of
tests of Fermi and Bose statistics is given in \cite{gr}.  The best bound
on the Fermi statistics of electrons is due to Ramberg and Snow\cite{ramb}.

{\bf Acknowledgements}

This work was supported in part by the National Science Foundation and by a
Semester Research Award from the University of Maryland, College Park.  It is
a pleasure to thank Alex Dragt for a careful reading of the manuscript and to
thank Joseph Sucher for trenchant criticism of an earlier version of this
paper.

\end{document}